\documentclass[twocolumn,prl,superscriptaddress,showpacs]{revtex4}

\usepackage{graphicx}
\usepackage{dcolumn}
\usepackage{multirow}
\usepackage{bm}
\usepackage{times}
\usepackage{color}
\usepackage{amstext}
\usepackage{amsmath}
\usepackage{amsfonts}
\usepackage[colorlinks,plainpages=false,linkcolor=black,urlcolor=blue,citecolor=black,pdfpagemode=UseNone,pdfstartview=FitH]{hyperref}
\begin{document}

\makeatletter
\long\def\@makecaption#1#2{%
  \par
  \vskip\abovecaptionskip
  \begingroup
   \small\rmfamily
   \sbox\@tempboxa{%
    \let\\\heading@cr
    #1 (color online). #2\hskip1pt%
   }%
   \@ifdim{\wd\@tempboxa >\hsize}{%
    \begingroup
     \samepage
     \flushing
     \let\footnote\@footnotemark@gobble
     #1 (color online). #2\hskip1pt\par
    \endgroup
   }{%
     \global \@minipagefalse
     \hb@xt@\hsize{\hfil\unhbox\@tempboxa\hfil}%
   }%
  \endgroup
  \vskip\belowcaptionskip
}%
\makeatother

\title{Optical conductivity of superconducting $\textrm{Rb}_2\textrm{Fe}_4\textrm{Se}_5$}

\author{A.~Charnukha}
\affiliation{Max-Planck-Institut f\"ur Festk\"orperforschung, Heisenbergstrasse 1, D-70569 Stuttgart, Germany}
\author{J.~Deisenhofer}
\affiliation{Experimental Physics V, Center for Electronic Correlations and Magnetism, Institute of Physics,
University of Augsburg, D-86159 Augsburg, Germany}
\author{D.~Pr\"opper}
\affiliation{Max-Planck-Institut f\"ur Festk\"orperforschung, Heisenbergstrasse 1, D-70569 Stuttgart, Germany}
\author{M.~Schmidt}
\author{Z.~Wang}
\affiliation{Experimental Physics V, Center for Electronic Correlations and Magnetism, Institute of Physics,
University of Augsburg, D-86159 Augsburg, Germany}
\author{Y.~Goncharov}
\affiliation{Experimental Physics V, Center for Electronic Correlations and Magnetism, Institute of Physics,
University of Augsburg, D-86159 Augsburg, Germany}
\affiliation{Institute of General Physics, Russian Academy of Sciences, 119991 Moscow, Russia}
\author{A.~N.~Yaresko}
\affiliation{Max-Planck-Institut f\"ur Festk\"orperforschung, Heisenbergstrasse 1, D-70569 Stuttgart, Germany}
\author{V.~Tsurkan}
\affiliation{Experimental Physics V, Center for Electronic Correlations and Magnetism, Institute of Physics,
University of Augsburg, D-86159 Augsburg, Germany}
\affiliation{Institute of Applied Physics, Academy of Sciences of Moldova, MD-2028 Chisinau, R. Moldova}
\author{B.~Keimer}
\affiliation{Max-Planck-Institut f\"ur Festk\"orperforschung, Heisenbergstrasse 1, D-70569 Stuttgart, Germany}
\author{A.~Loidl}
\affiliation{Experimental Physics V, Center for Electronic Correlations and Magnetism, Institute of Physics,
University of Augsburg, D-86159 Augsburg, Germany}
\author{A.~V.~Boris}
\affiliation{Max-Planck-Institut f\"ur Festk\"orperforschung, Heisenbergstrasse 1, D-70569 Stuttgart, Germany}

\begin{abstract}
We report the complex dielectric function of high-quality nearly-stoichiometric $\textrm{Rb}_2\textrm{Fe}_4\textrm{Se}_5$ (RFS) single crystals with $T_{\mathrm{c}}=32\ \textrm{K}$ determined by wide-band spectroscopic ellipsometry and time-domain transmission spectroscopy in the spectral range $1\ \textrm{meV}\leq\hbar\omega\leq6.5\ \textrm{eV}$ at temperatures $4\ \textrm{K}\leq T\leq300\ \textrm{K}$. This compound simultaneously displays a superconducting and a semiconducting optical response. It reveals a direct band-gap of $\approx0.45\ \textrm{eV}$ determined by a set of spin-controlled interband transitions. Below $100\ \textrm{K}$ we observe in the lowest THz spectral range a clear metallic response characterized by the negative dielectric permittivity $\varepsilon_1$ and bare (unscreened) $\omega_{\textrm{pl}}\approx100\ \textrm{meV}$. At the superconducting transition this metallic response exhibits a signature of a superconducting gap below $8\ \textrm{meV}$. Our findings suggest a coexistence of superconductivity and magnetism in this compound as two separate phases.
\end{abstract}

\pacs{74.25.Gz,74.70.Xa,71.15.Mb}

\maketitle
\par In the family of iron-pnictide/chalcogenide superconductors most research effort has so far been applied to the so-called 122 compounds with Fe-As conducting planes due to the high quality and large size of the single crystals available. They bear all hallmarks of this new class of superconductors such as an itinerant antiferromagnetic ground state of the parent compounds, multiple bands crossing the Fermi level, superconducting transition temperatures up to 40~K, and a resonance peak in the inelastic neutron scattering signal at a $(0,\pi)$ or $(\pi,0)$ $\textbf{k}$-vector in the superconducting state~\cite{Johnston_Review_2010}, suggesting novel superconductivity with $s$-wave symmetry and a sign change of the order parameter between the hole and electron Fermi pockets~\cite{Mazin_NatureInsights_2010}. Throughout the phase diagram these compounds are metals with a plasma frequency $\omega_{\textrm{pl}}\approx1.6\ \textrm{eV}$~\cite{PhysRevLett.101.257005,CharnukhaNatCommun2011,PhysRevB.82.054518}. In the superconducting state the corresponding optical conductivity is fully suppressed below $2\Delta$ due to the formation of a superconducting condensate with a London penetration depth of $\lambda_{\textrm{L}}\approx220\ \textrm{nm}$~\cite{PhysRevLett.101.107004,PhysRevB.81.214508,2011arXiv1103.0938C}.

\begin{figure}[!tb]
\includegraphics[width=3.4in]{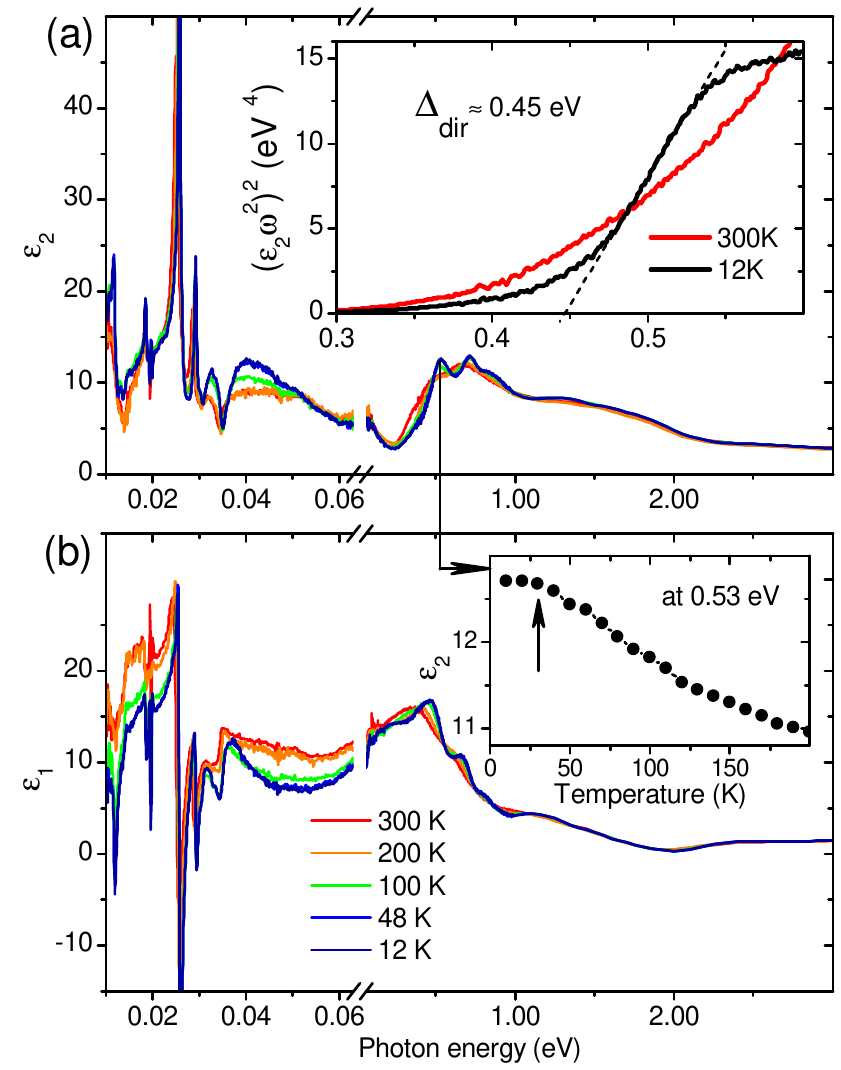}
\caption{\label{fig:firexperimentelli} (a) Imaginary and (b) real parts of the dielectric function of RFS in the $0.01-6.5\ \textrm{eV}$ spectral range at different temperatures. (Inset in (a)) Plot of $(\varepsilon_2(\omega)\omega^2)^2$ near the absorption edge. The intersection of the dashed line with the energy axis defines the direct energy gap $\Delta_{\textrm{dir}}=0.45\ \textrm{eV}$ at $12\ \textrm{K}$. (Inset in (b)) Temperature dependence of $\epsilon_2(0.53\ \textrm{eV})$.}
\end{figure}

Recently, iron selenide compounds have been synthesized in this class of superconductors~\cite{PhysRevB.82.180520,PhysRevB.83.212502,APL10.10631.3549702,PhysRevB.83.060512}. They were first believed to crystallize in the same I4/mmm symmetry of $\textrm{ThCr}_2\textrm{Si}_2$ type but soon it became clear that there is an inherent iron-deficiency order present in these materials with a chiral $\sqrt{5}\times\sqrt{5}\times1$ superstructure, which reduces the symmetry to I4/m and makes it more appropriate to classify these materials into the 245 stoichiometry~\cite{C1SC00070E}. The Fe-defect and antiferromagnetic orders occur at rather high transition temperatures of $400-550\ \textrm{K}$. Neutron-scattering studies showed that these compounds possess a magnetic moment on iron atoms of about $3.3\ \mu_{\textrm{B}}$~\cite{2011arXiv1102.0830B}, which is unusually large for iron pnictides. At the same time a resonance peak has been observed by the inelastic neutron scattering below $T_{\textrm{c}}\approx32~\textrm{K}$ at an energy of $\hbar\omega_{\textrm{res}}=14\ \textrm{meV}$ and the $\mathbf{k}$-vector $(0.5,0.25,0.5)$ in the unfolded Fe-sublattice notation~\cite{2011arXiv1107.1703P}, which is also unprecedented for the iron pnictides. It is still under debate how superconductivity with such a high transition temperature can exist on such a strong magnetic background although there are some indications of an inherent phase separation in iron chalcogenides~\cite{2011arXiv1107.0412R,2011arXiv1102.1381Y,2011arXiv1108.0069L,2011arXiv1108.2895W}. A further complication arises from the fact that, unlike their 122 counterparts, 245 iron selenides show a semiconducting optical response and no free charge carrier conductivity has been reported so far~\cite{PhysRevB.83.220507}. 

In this work we provide first direct evidence for a free charge carrier contribution to the optical response in RFS. We show that the charge-carrier density is small with $\omega_{\textrm{pl}}\approx100\ \textrm{meV}$. This metallic response experiences a weak modification upon cooling into the superconducting state. This evidence together with the results of resistivity, magnetization, specific-heat measurements~\cite{2011arXiv1107.3932T}, and M\"ossbauer spectroscopy~\cite{2011arXiv1108.3006K} indicate that in this compound superconductivity and magnetism coexist as two separate phases.

The optimally-doped RFS single crystals were grown by the Bridgman method (batch BR16 in Ref.~\onlinecite{2011arXiv1107.3932T}). From DC resistivity, magnetization and specific-heat measurements we obtained $T_{\textrm{c}}\approx32\ \textrm{K}$. Sample cleaving and handling was carried out in argon atmosphere at all times prior to every optical measurement. The complex dielectric function $\varepsilon(\omega)=\varepsilon_1(\omega)+i\varepsilon_2(\omega)=1+4\pi i\sigma(\omega)/\omega$, where $\sigma(\omega)$ is the complex optical conductivity, was obtained in the range $0.01-6.5\ \textrm{eV}$ using broadband ellipsometry, as described in Ref.~\onlinecite{boris:027001}, in a combination with time-domain THz spectroscopy in the $1-10\ \textrm{meV}$ spectral range (TPS spectra 3000, TeraView Ltd.). The far-infrared optical response was measured at the infrared beamline of the ANKA synchrotron light source at Karlsruhe Institute of Technology, Germany.

\begin{figure}[!tb]
\includegraphics[width=3.4in]{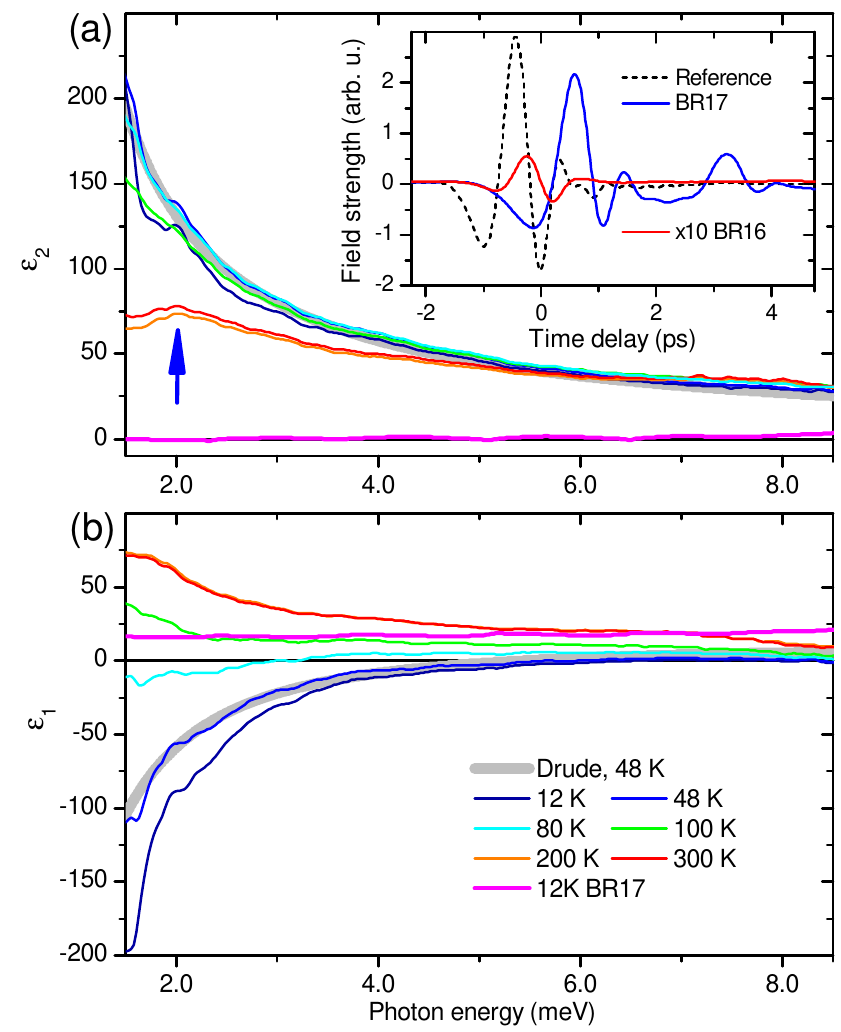}
\caption{\label{fig:firexperimenttd} (a) Imaginary and (b) real parts of the dielectric function of RFS in the THz spectral region obtained via the time-domain transmission spectroscopy. Blue arrow marks a low-energy electronic excitation. Thick gray lines show a Drude fit to $\varepsilon_1(\omega)$ and $\varepsilon_2(\omega)$ as described in the text. (Inset) Electric field transients transmitted through a $25\ \mu\textrm{m}$-thick superconducting sample BR16 (scaled by 10, red line) and $100\ \mu\textrm{m}$-thick transparent insulating sample BR17 (blue line). The dashed line plots the reference signal.}
\end{figure}

The imaginary and real parts of the complex dielectric function in the $0.01-6.5\ \textrm{eV}$ spectral range are shown in Figs.~\ref{fig:firexperimentelli}~(a) and~(b), respectively. Down to $10\ \textrm{meV}$ the sample does not reveal any metallic behavior as is evident from $\varepsilon_1$, which remains positive at all temperatures. It also displays several infrared-active optical phonons, similar to previously observed in the far-infrared optical response of a semiconducting $\textrm{K}_2\textrm{Fe}_4\textrm{Se}_5$ (KFS)~\cite{PhysRevB.83.220507}. More infrared-active phonons observed than allowed by tetragonal symmetry of the 122 unit cell supports the reduction of the Brillouin zone due to the ordering of iron vacancies. Throughout the whole far-infrared $10-100\ \textrm{meV}$ spectral range we find a rapid increase of the electronic background in $\sigma_1(\omega)$ between $200$ and $100\ \textrm{K}$ with a concomitant decrease in $\varepsilon_1$. At higher frequencies the optical response features the onset of interband transitions around $0.25\ \textrm{eV}$. It is followed by an absorption edge at $\approx0.45\ \textrm{eV}$ formed by direct interband transitions, as shown in the inset of Fig.~\ref{fig:firexperimentelli}~(a). Unlike in 122 compounds~\cite{PhysRevLett.101.257005}, the lowest-lying absorption band peaked at about $0.6\ \textrm{eV}$ reveals three separate contributions at low temperatures similar to the two contributions reported for the semiconducting KFS~\cite{PhysRevB.83.220507}. The inset in Fig.~\ref{fig:firexperimentelli}~(b) shows the strong temperature dependence of one of these absorption bands in the magnetic state, which is linear in a broad temperature range. This behavior is fully consistent with the temperature dependence of the magnetic Bragg peak intensity, including the saturation at $30-40\ \textrm{K}$~\cite{2011arXiv1102.2882Y}, which suggests a spin-controlled character of these interband transitions.

As the compound remains transparent down to $10\ \textrm{meV}$, a pathway opens to apply time-domain transmission spectroscopy to obtain the complex dielectric function of RFS also at THz frequencies, should sufficiently thin samples be obtained. It was indeed possible to achieve sizable transmission in cleaved flakes of RFS about $25\ \mu\textrm{m}$ thick. Figures~\ref{fig:firexperimenttd}~(a) and~(b) show $\varepsilon_2$ and $\varepsilon_1$, respectively, obtained in the $1-10\ \textrm{meV}$ spectral range using this technique for temperatures $4\leq T\leq300\ \textrm{K}$, along with the optical response of a $100\ \mu\textrm{m}$-thick insulating sample BR17 grown by the same method~\cite{2011arXiv1107.3932T}. The latter shows a typical frequency independent insulating response even at $12\ \textrm{K}$ (magenta lines in Fig.~\ref{fig:firexperimenttd}). Typical electric-field transients obtained on these samples are shown in the inset. The superconducting BR16 shows much stronger absorption (solid red line) than the insulating BR17 (solid blue line) due to a high level of the electronic background in $\varepsilon_2(\omega)$. At room temperature and down to $100\ \textrm{K}$ BR16 remains semiconducting, with $\varepsilon_1(\omega)$ positive in the whole spectral range. However, unlike the insulating BR17 (thick magenta line in Fig.~\ref{fig:firexperimenttd}~(b)), it exhibits an upturn at lowest energies, which indicates a low-energy electronic mode peaked at $2\ \textrm{meV}$ (blue arrow in Fig.~\ref{fig:firexperimenttd}~(a)). Below $100\ \textrm{K}$ a clear metallic response with negative $\varepsilon_1(\omega)$ rapidly develops. Already at $80\ \textrm{K}$ the zero-crossing in $\varepsilon_1(\omega)$ corresponds to a screened plasma frequency of $3\ \textrm{meV}$, which reaches $6.5\ \textrm{meV}$ as the temperature is lowered further. The metallic response can be fitted by two Drude terms with $\omega_{\textrm{pl,1}}\approx20\ \textrm{meV}$, $\gamma_1\approx1\ \textrm{meV}$ and $\omega_{\textrm{pl,2}}\approx95\ \textrm{meV}$, $\gamma_2\approx40\ \textrm{meV}$ for the plasma frequencies and scattering rates of the narrow and broad components at $48\ \textrm{K}$, respectively (thick gray lines in Fig.~\ref{fig:firexperimenttd}). The total charge-carrier density is given by $\omega_{\textrm{pl}}=\sqrt{\omega_{\textrm{pl,1}}^2+\omega_{\textrm{pl,2}}^2}\approx100\ \textrm{meV}$. The observed crossover from semiconducting to metallic behavior with decreasing temperature below $100\ \textrm{K}$ is in full agreement with the temperature dependence of the DC resistivity~\cite{2011arXiv1107.3932T}. The spectral weight of the low-energy electronic mode in the semiconducting state amounts to about $4\%$ of the total spectral weight of the free charge carriers and might originate from the narrow Drude component. This low-energy mode can represent a collective electronic excitation pinned by structural defects like iron vacancies.

\begin{figure}[!t]
\includegraphics[width=3.4in]{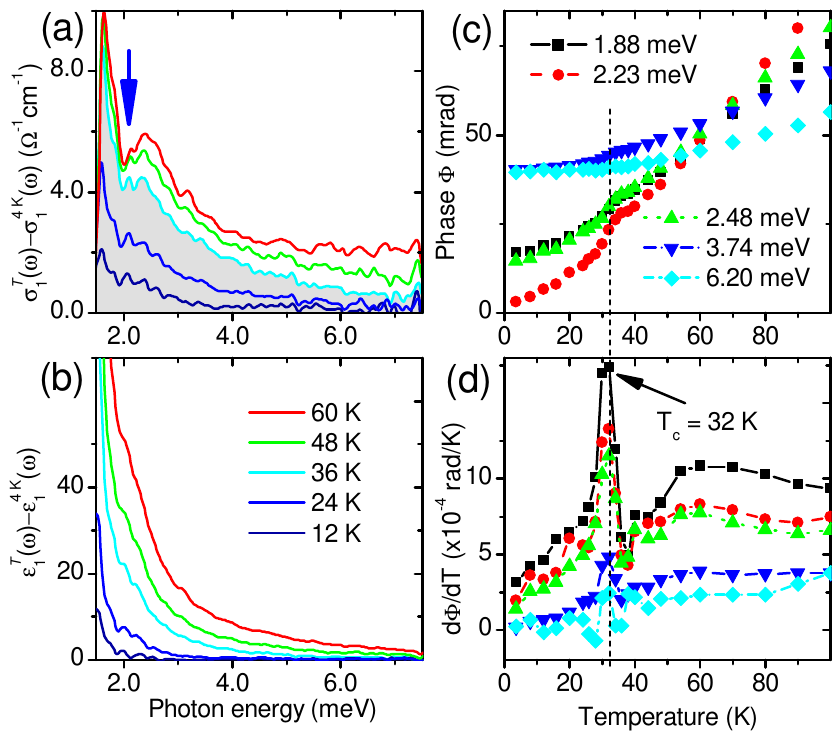}
\caption{\label{fig:tdsc} Difference spectra (a) $\sigma_1(T,\omega)-\sigma_1(4~\textrm{K},\omega)$ and (b) $\varepsilon_1(T,\omega)-\varepsilon_1(4~\textrm{K},\omega)$ of a $25~\mu\textrm{m}$ superconducting sample. Blue arrow marks the same low-energy electronic excitation as in Fig.~\ref{fig:firexperimenttd}~(a). Shaded area indicates the spectral weight used for the estimation of the $\omega_{\textrm{pl}}$ of the superconducting condensate (see text). (c) Temperature dependence of the time-domain transmission phase and (d) its temperature derivative for several frequencies. Superconducting transition temperature of 32K (dashed line).}
\end{figure}

The discovery of itinerant charge carriers in the optimally-doped RFS requires a more detailed study of its low-temperature optical response in the vicinity of the superconducting transition temperature. Difference spectra of the real parts of the optical conductivity and the dielectric function are shown in Fig.~\ref{fig:tdsc}~(a) and~(b), respectively, for several temperatures between $12$ and $60\ \textrm{K}$. The sample shows moderate changes in the normal state and a rapid decrease of the optical conductivity below $T_{\textrm{c}}$. The missing area between $36$ and $4\ \textrm{K}$ shown in grey in Fig.~\ref{fig:tdsc}~(a) gives rise to a characteristic $-1/\omega^2$ contribution to $\varepsilon_1(\omega)$. Using a Kramers-Kronig consistency analysis of the relative changes in the complex dielectric function (see Ref.~\onlinecite{CharnukhaNatCommun2011}), we determine the superconducting plasma frequency $\omega^{\textrm{SC}}_{\textrm{pl}}\approx10\ \textrm{meV}$. Further evidence for the superconductivity-induced nature of these changes in the optical response comes from the temperature dependence of the transmission phase shown in Fig.~\ref{fig:tdsc}~(c) for several frequencies. At all frequencies up to $8\ \textrm{meV}$ there is a kink at $T_{\textrm{c}}\approx32\ \textrm{K}$, which gets progressively smaller as the frequency increases. This effect is more obvious in the temperature derivative of the transmission phase shown for the same frequencies in Fig.~\ref{fig:tdsc}~(d). In addition, we observe a double-peak structure around $2\ \textrm{meV}$ (blue arrow in Fig.~\ref{fig:tdsc}~(a); see also Fig.~\ref{fig:firexperimenttd}~(a)), which is overwhelmed by the electronic background at $300\ \textrm{K}$ but clearly stands out at lower temperatures due to reduced electron scattering and even persists in the superconducting state. This feature might have its origin in the electron and hole bound states induced by iron vacancies recently observed in an STM/STS study on KFS in Ref.~\onlinecite{2011arXiv1108.0069L}, which might serve as pinning centers for a collective electronic excitation. 
  
\begin{figure}[!tb]
\includegraphics[width=3.4in]{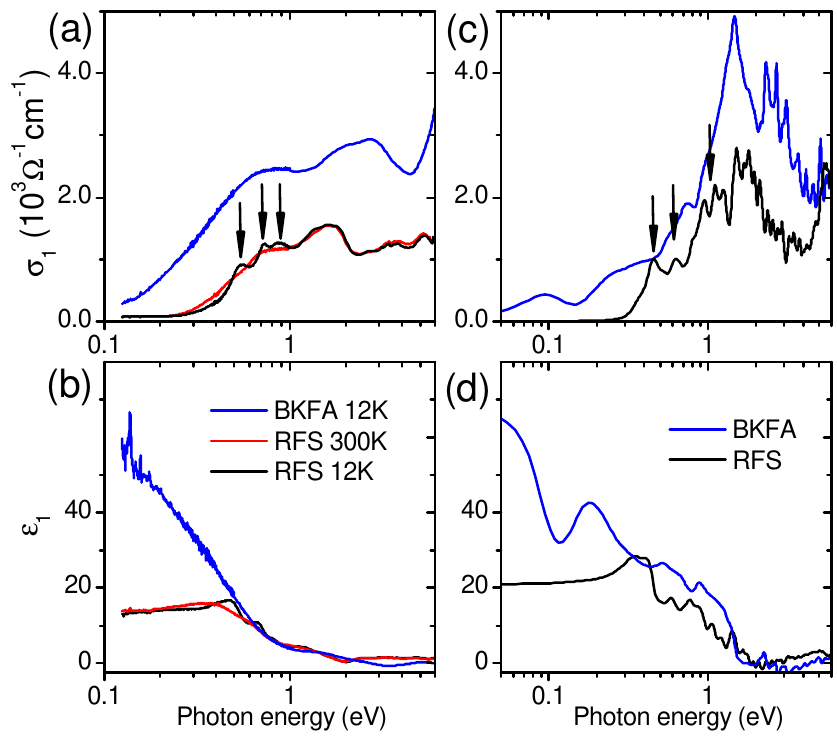}
\caption{\label{fig:lda_sigma} (Left panels) Comparison of the experimentally-obtained real parts of the (a) optical conductivity and (b) dielectric function of an optimally-doped BKFA at 12K and RFS at $300$ and $12~\textrm{K}$. (Right panels) Same as in left panels as obtained in LDA calculations. The arrows in (a) and (c) indicate the narrow low-energy optical bands that can be resolved in RFS at low temperatures but not in BKFA.}
\end{figure}

It is known that LDA calculations provide an adequate description of the band structure and optical conductivity of iron pnictides~\cite{CharnukhaNatCommun2011,PhysRevLett.103.026404,PhysRevLett.104.057008} as long as a moderate mass and bandwidth renormalization is taken into account. We compare the experimentally obtained $\sigma_1(\omega)$ with the theoretical prediction for the $\textrm{Rb}_2\textrm{Fe}_4\textrm{Se}_5$ compound. The calculation was performed for the experimental $\sqrt{5}\times\sqrt{5}\times1$ superstructure~\cite{C1SC00070E} assuming the so-called block-checkerboard
antiferromagnetic order of Fe moments~\cite{PhysRevB.83.233205}. In Figs.~\ref{fig:lda_sigma}~(a) and~(b) the experimentally obtained spectra of $\sigma_1(\omega)$ and $\varepsilon_1(\omega)$ for the 122 BKFA (optimally-doped $\textrm{Ba}_{0.68}\textrm{K}_{0.32}\textrm{Fe}_2\textrm{As}_2$) and 245 RFS systems are compared to the results of LDA calculations (c)~and~(d), respectively. Already a direct comparison of the experimentally-obtained $\sigma_1(\omega)$ and $\varepsilon_1(\omega)$ of BKFA and RFS show that the overall structure of the interband transitions in these two classes of superconductors is very similar apart from frequency shifts (blue and red lines in Fig.~\ref{fig:lda_sigma}~(a), respectively; the overall shape of the interband optical conductivity of BKFA is virtually unchanged between $300$ and $10\ \textrm{K}$, see supplementary material in Ref.~\onlinecite{2011arXiv1103.0938C}). The most striking difference is the narrowing of the absorption bands in RFS around $0.6\ \textrm{eV}$ at low temperatures uncovering three distinct components (black arrows in Fig.~\ref{fig:lda_sigma}~(a)). Similarly, appearance of this fine structure can be observed in the LDA calculations, as shown in Fig.~\ref{fig:lda_sigma}~(c) for BKFA and RFS (blue and black lines, respectively, arrows indicate three possible contributions to the resolved fine structure of the $0.6\ \textrm{eV}$ absorption band). It has already been shown in Ref.~\onlinecite{2011arXiv1103.0938C} by subtracting the itinerant contribution to the infrared optical response of BKFA that the low-frequency dielectric permittivity in this compound has an anomalously large value of about $60$ most likely due to the high polarizability of the Fe-As bond. In RFS, the very weak and narrow free charge carrier response allows for determination of the low-frequency permittivity already in the raw data. Figure~\ref{fig:lda_sigma}~(b) compares $\varepsilon_1(\omega)$ for BKFA at $12\ \textrm{K}$ (with the itinerant response subtracted~\cite{2011arXiv1103.0938C}) and RFS at $12$ and~$300\ \textrm{K}$ (raw data). It is clear that the low-energy permittivity of RFS is about three times smaller than that of BKFA. This trend is very well reproduced in our LDA calculations shown in Fig.~\ref{fig:lda_sigma}~(d), in which a comparable decrease is obtained.

Our LDA calculations show, consistent with previous work~\cite{PhysRevB.83.233205,PhysRevLett.107.056401}, that the stoichiometric 245 compounds are semiconducting and minor doping of either sign results in a very complicated Fermi surface in the magnetic state with only one type of carriers present. This is fully consistent with the weak metallic response of a nearly stoichiometric RFS observed in this work as well as with the overall phase diagram reported for this system in Ref.~\onlinecite{2011arXiv1107.3932T}, where a narrow doping range was found for the superconducting phase bounded by an insulating and a semiconducting phase on the underdoped and overdoped sides, respectively. In the same Ref.~\onlinecite{2011arXiv1107.3932T} it is shown that the electronic specific-heat exhibits a rather small superconductivity-induced anomaly at the superconducting transition temperature. Together with the small effect of superconductivity on the itinerant optical response observed here, it implies that superconductivity in RFS is not a uniformly bulk phenomenon. This conclusion is also consistent with a practically doping-independent superconducting transition temperature observed in Ref.~\onlinecite{2011arXiv1107.3932T} assuming that the superconducting phase stabilizes at the same doping level, while the excess is doped into the coexisting phase(s).

In summary, we obtained the complex dielectric function of a $\textrm{Rb}_2\textrm{Fe}_4\textrm{Se}_5$ superconductor with $T_{\textrm{c}}\approx32\ \textrm{K}$ in the spectral range from $1\ \textrm{meV}$ to $6.5\ \textrm{eV}$. Comparison with our LDA calculations shows that the optical response of this material is well reproduced and is close to its Fe-As based counterparts in the 122 family. Strikingly, unlike in iron pnictides, the absorption band at $0.6\ \textrm{eV}$ experiences a spin-controlled narrowing into three sub-bands in the magnetic state. We further demonstrated that the superconducting RFS displays a clear metallic response in the THz spectral range below $100\ \textrm{K}$ with $\omega_{\textrm{pl}}\approx100\ \textrm{meV}$, which can be divided into a narrow and a broad component and is partially suppressed in the superconducting state giving rise to a superconducting condensate with a plasma frequency of $\omega_{\textrm{pl}}^{\textrm{SC}}\approx10\ \textrm{meV}$. Such a small charge-carrier density suggests that the optical conductivity of the superconducting RFS represents an effective-medium response of two separate phases dominated by the magnetic semiconducting phase.

This project was supported by the German Science Foundation under grants BO 3537/1-1 and DE1762/1-1 within SPP 1458, as well as TRR80 (Augsburg-Munich). We gratefully acknowledge Y.-L.Mathis for support at the infrared beamline of the synchrotron facility ANKA at the Karlsruhe Institute of Technology.


\end{document}